\documentclass[11pt,preprint]{aastex}

\shorttitle{Transformation in HK catalog}
\shortauthors{C. Zhao \& H.J. Newberg}

\begin{document}

\title{Transformation From SDSS Photometric System to Johnson-Morgan-Cousins System in HK Survey}

\author{Chongshan Zhao \& Heidi Jo Newberg}
\affil{Department of Physics, Applied Physics \& Astronomy, \\
    Rensselaer Polytechnic Institute, Troy, NY 12180}
\email{newbeh@rpi.edu}

\altaffiltext{1}{}

\begin{abstract}
We calculate the transformation from the Sloan Digital Sky Survey (SDSS) 
photometric system to the Johnson-Morgan-Cousins System in the HK Survey.  
This research was done in late 2001, so the SDSS photometry was taken from 
the databases prior to the release of DR1.  This paper is being posted 
because it is referenced in other papers in the literature, but will not 
be submitted to a refereed journal because it uses unpublished versions of
the catalogs.
\end{abstract}

%\keywords{stellar color: UBV system; SDSS photometry; HK survey}

\section{Introduction}
The Sloan Digital Sky Survey (SDSS) photometric catalogs are an important source
of stellar photometry, and must be understood in the context of decades of
stellar research using different filter standards.  In this paper a transformation
is computed between the available SDSS photometry in October 2001 and stars of the
HK objective prism survey as provided by T. Beers, private communication.  A 
description of the HK survey and an earlier version of the catalog can be found at 
\citet{psb91}.  The overlaps between the catalogs are relatively few because the 
faint limit of the HK survey is at approximately the same magnitude as the saturation 
limit of the SDSS.  Neither of the catalogs compared here are standard versions,
but were the only comparables available at the time.  Technical SDSS details can be
found in \citet{getal98}; \citet{hetal01}; \citet{petal03}; \citet{setal02}; and 
\citet{yetal00}.  More recent
transformation equations can be obtained from \citet{jetal05}; \citet{kbt05}; 
\citet{bkt05}; and \citet{wwh05};
and two other unpublished determinations that can be found in the documentation
for the SDSS DR4 at:

{\it http://www.sdss.org/dr4/algorithms/sdssUBVRITransform.html }.

\section{Query Consideration and Data Reduction}
Table 1 lists the stars that were common to the HK survey and the SDSS survey in October
2001.  Generation of both catalogs was a work in progress at that time.
We selected from the HK catalog all stars whose J2000 coordinates matched
SDSS catalog entries within:
\begin{eqnarray}
\Delta(RA)<7."2 (0.002^{\circ})& \Delta(Dec)<7."2
\end{eqnarray}
We rejected any matches in which the SDSS star was saturated by checking the catalog
flags:
\begin{equation}
(objFlags\,\, \& \,\,OBJECT\_SATUR) == 0 
\end{equation}
We use the magnitude calculated from a fit of modeled stellar profile
(point-spread-function, or PSF magnitude) to each object. 
Because these stars are too bright for sky noise affect the quality
of the photometry, the use of aperture magnitudes would have made little 
difference.

All photometry in Table 1 has been corrected for interstellar reddening
using the $E_{B-V}$ determined from the HK survey.  Corrections for the
SDSS photometry are determined from the standard extinction curve of
\citet{ccm89}, which for SDSS filters yields:
\begin{eqnarray}
A_{u^*}=5.2E_{B-V}; & A_{g^*}=3.2E_{B-V};& A_{r^*}=2.8E_{B-V}; \nonumber\\
A_{i^*}=2.1E_{B-V}; & A_{z^*}=1.5E_{B-V} \nonumber
\end{eqnarray}
We noticed that the value of $(U-B)$ is -9.990 for some stars in the original HK list. 
They were replaced by "$\cdots$" in the Table 1. When we calculate the 
transformation which requires $U$ band, these stars were not included.
Three stars were removed from the original matched list because
their SDSS photometry was suspicious: (RA, dec) = 
{\bf (199.9265, 3.9129)}, {\bf (224.7835, -0.2537)}, {\bf (200.7934, 4.6075)}.
Two of the stars were close to other bright starts which may have confused the
SDSS object deldender.

\begin{figure}
\plotone{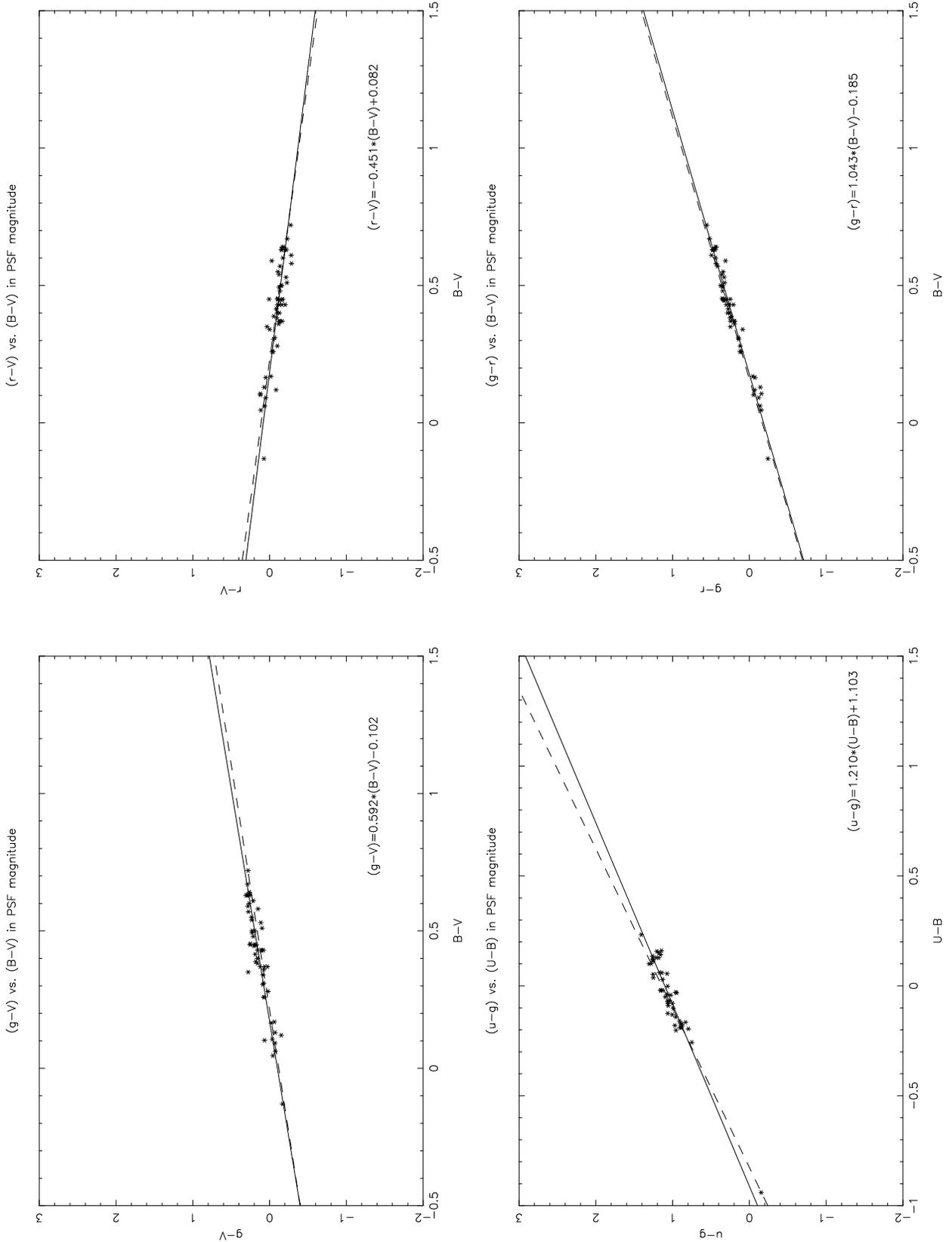}
\caption{Color-Color plots of UBV system against SDSS PSF magnitude.  The solid
line is the least squares fit to the data.  The dashed line indicates the
transformation equation of Fukugita et al. (1996).}
\end{figure}

\section{Comparison of derived transformation to Fukugita et al. 1996}

Figure 1 shows the color-color plots for the SDSS PSF magnitude against the HK
survey U-B-V magnitudes of these stars. The coeffices in the equations, plotted as the lines, 
are derived using the {\bf\it Method of Least Squares}. We notice that one star, 
{\bf (221.5996, -0.1158)} is half a magnitude brighter in SDSS filters than we expect
(though the colors are the same). It is labeled using a bold font in Table 1. 
There is no reason to remove it from the catalog but we ignore 
it in all fits for filter transformations.

\citet{fetal96} described SDSS photometric system and gave 
the approximate color transformation equations from the 
Johnson-Morgan-Cousins system to the SDSS system.  A comparison of our
transformations to theirs is given below:

\begin{eqnarray*}
   Fukugita's\, paper  &  & Our Result \nonumber\\
g^*=V+0.56(B-V)-0.12 & & g^*=V+0.592(B-V)-0.102 \nonumber\\
r^*=V-0.49(B-V)+0.11 & & r^*=V-0.451(B-V)+0.082 \nonumber\\
u^*-g^*=1.38(U-B)+1.14 & & u^*-g^*=1.210(U-B)+1.103  \nonumber\\
g^*-r^*=1.05(B-V)+1.14 & & g^*-r^*=1.043(B-V)-0.185  \nonumber
\end{eqnarray*}

For the blue stars from which our transformation was derived, our transformations 
are similar to Fukugita's.  The ($u^*-g^*$) transformation is the only one that
is significantly discrepant.  This is because the
actual $u^*$ filter response is different from the theoretical curve 
used by Fukugita et al.(1996).

\section{Inverse Transformation Equations}

We will discuss the inverse transformation from SDSS PSF magnitude to UBV system 
in this section.  Magnitudes in the $g^*$ filter and $g^*-r^*$ color are used as the 
primary parameters in the equations since the noise in $u^*$ is typically higher. 
We also considered other combinations of filters. 
Figures 2 and 3 show the results; the transformation is summaried
below: 
\begin{figure}
\plotone{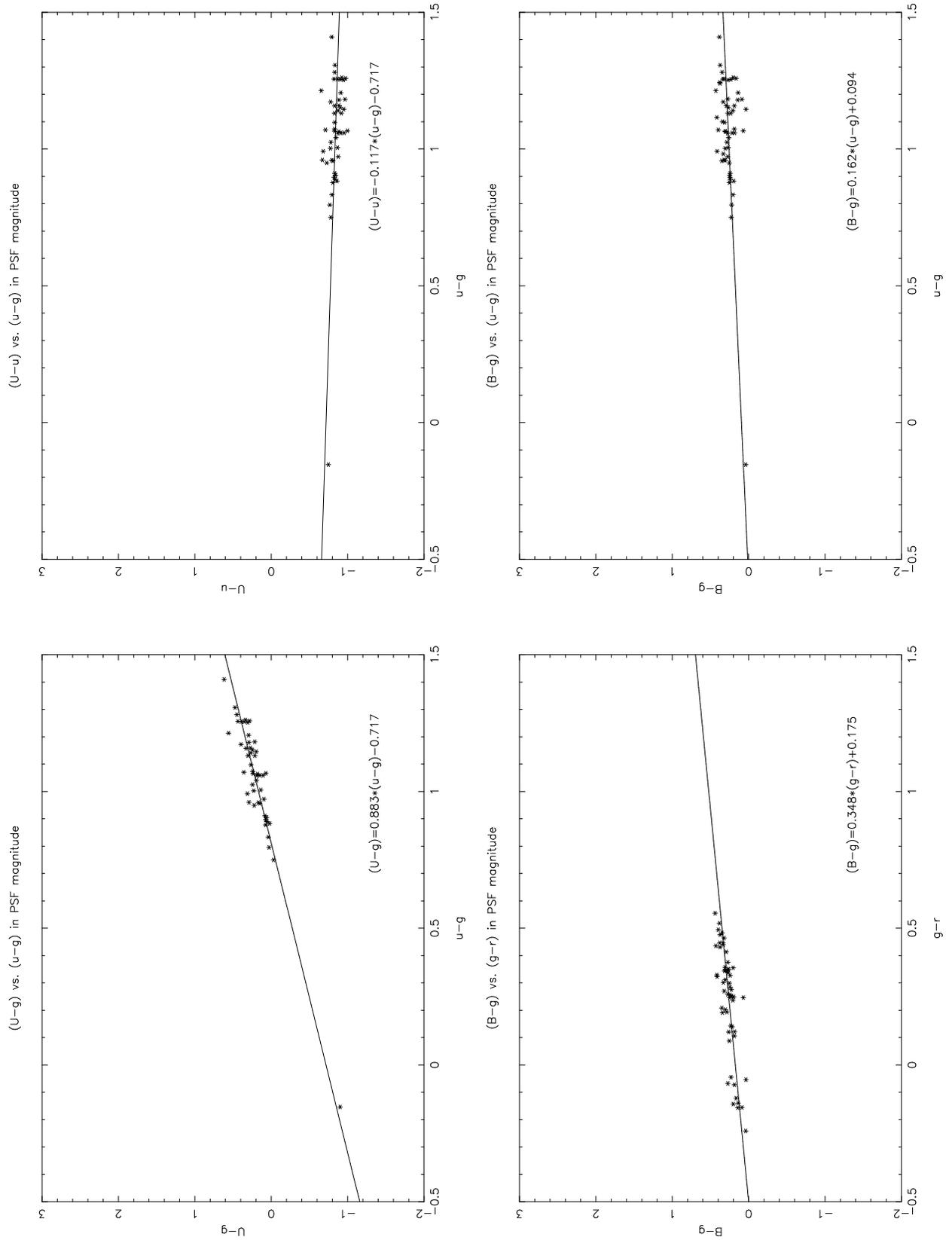}
\caption{Transformation plot from SDSS photometric system to UBV system.}
\end{figure}

\begin{figure}
\plotone{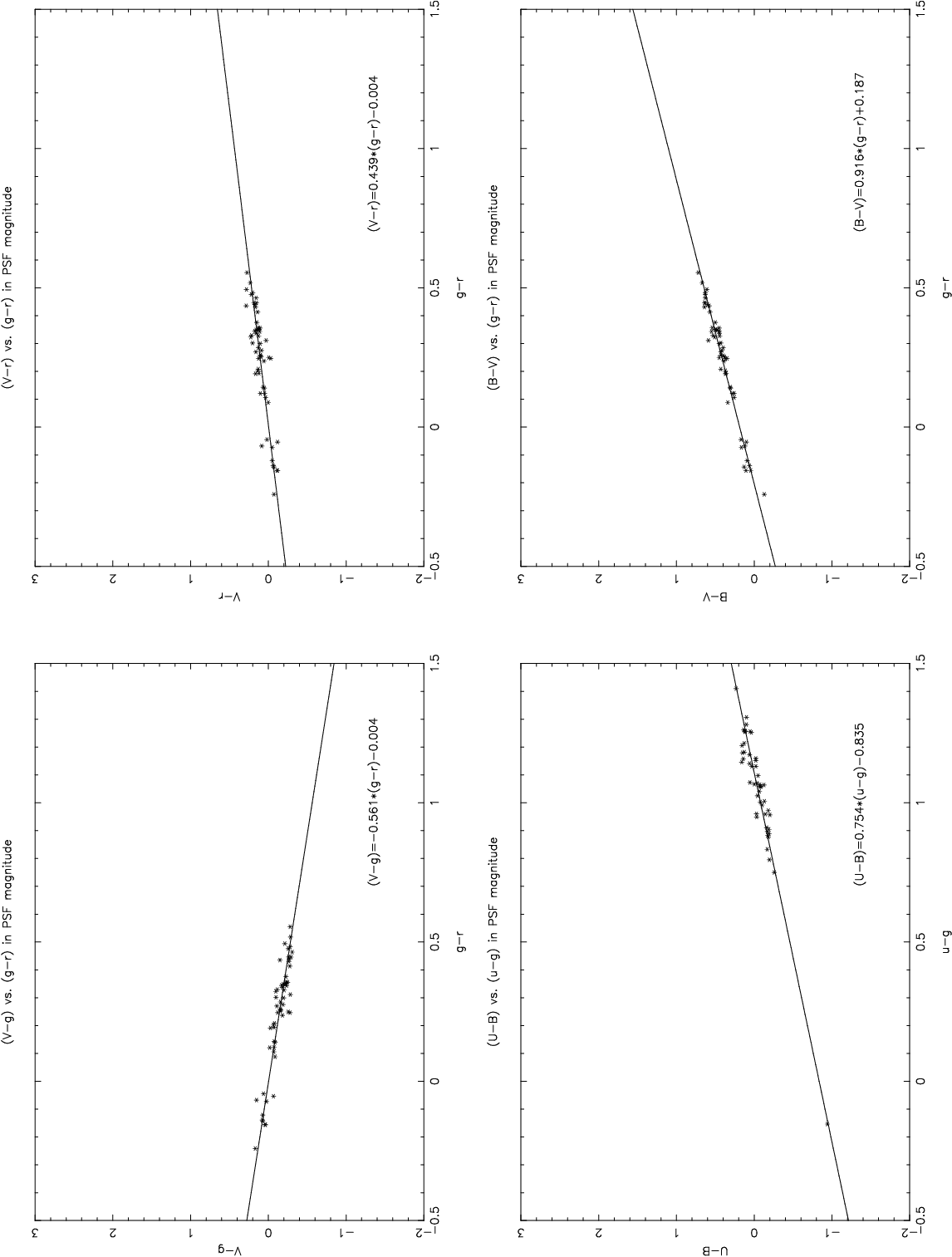}
\caption{Transformation plot from SDSS photometric system to UBV system.}
\end{figure}

\begin{eqnarray*}
U=g^*+0.883(u^*-g^*)-0.717 & & U=u^*-0.117(u^*-g^*)-0.717 \nonumber\\
B=g^*+0.348(g^*-r^*)+0.175 & & B=g^*+0.162(u^*-g^*)+0.094 \nonumber\\
V=g^*-0.561(g^*-r^*)-0.004 & & V=r^*+0.439(g^*-r^*)-0.004 \nonumber\\
(U-B)=0.754(u^*-g^*)-0.835  \nonumber \\
(B-V)=0.916(g^*-r^*)+0.187  \nonumber
\end{eqnarray*}

\section{Conclusion}

We have derived transformation equations between HK catalog photometry and
SDSS photometry.

\acknowledgments

Funding for the SDSS and SDSS-II has been provided by the Alfred P. Sloan Foundation, the Participating Institutions, the National Science Foundation, the U.S. Department of Energy, the National Aeronautics and Space Administration, the Japanese Monbukagakusho, the Max Planck Society, and the Higher Education Funding Council for England. The SDSS Web Site is http://www.sdss.org/.

The SDSS is managed by the Astrophysical Research Consortium for the Participating Institutions. The Participating Institutions are the American Museum of Natural History, Astrophysical Institute Potsdam, University of Basel, Cambridge University, Case Western Reserve University, University of Chicago, Drexel University, Fermilab, the Institute for Advanced Study, the Japan Participation Group, Johns Hopkins University, the Joint Institute for Nuclear Astrophysics, the Kavli Institute for Particle Astrophysics and Cosmology, the Korean Scientist Group, the Chinese Academy of Sciences (LAMOST), Los Alamos National Laboratory, the Max-Planck-Institute for Astronomy (MPIA), the Max-Planck-Institute for Astrophysics (MPA), New Mexico State University, Ohio State University, University of Pittsburgh, University of Portsmouth, Princeton University, the United States Naval Observatory, and the University of Washington.

\begin{deluxetable}{rccccccccccccc}
\rotate
\tablewidth{0pc}
\
\tablehead{
& HK Coord &\multicolumn{2}{c}{GSC Coords} & \multicolumn{2}{c}{SDSS Coords} &\multicolumn{3}{c}{UBV system } &
\multicolumn{5}{c}{SDSS PSF magnitude}\\
& & RA & Dec&  RA & Dec&  U & B & V & $u^{*}$ & $g^{*}$ & $r^{*}$ & $i^{*}$ & $z^{*}$
}
\tablecaption{Catalog of HK Survey Cross-correlated with SDSS non-EDR Database}
\startdata

        1 & 303110077 & 198.6167 &  3.4511 & 198.6159 &  3.4523 &  15.591 &  15.358 &  14.688 &  16.384 &  14.974 &  14.456 &  14.305 &  14.252 \\
        2 & 303110073 & 198.6225 &  5.5703 & 198.6217 &  5.5714 &  14.930 &  14.950 &  14.640 &  15.847 &  14.716 &  14.573 &  14.572 &  14.604 \\
        3 & 303110061 & 198.7546 &  3.4669 & 198.7540 &  3.4681 &  15.451 &  15.338 &  14.708 &  16.271 &  15.015 &  14.551 &  14.404 &  14.289 \\
        4 & 303110056 & 199.4292 &  2.9122 & 199.4284 &  2.9136 &  15.090 &  15.180 &  14.730 &  15.960 &  14.902 &  14.557 &  14.487 &  14.475 \\
        5 & 303110040 & 199.7233 &  2.5875 & 199.7230 &  2.5878 &  15.320 &  15.340 &  14.770 &  16.205 &  15.046 &  14.633 &  14.534 &  14.485 \\
        6 & 303110048 & 199.8633 &  5.8553 & 199.8626 &  5.8556 &  15.160 &  15.350 &  14.950 &  15.997 &  15.107 &  14.851 &  14.776 &  14.767 \\
        7 & 303110046 & 199.9279 &  3.9136 & 199.9281 &  3.9145 &  15.401 &  15.348 &  14.718 &  16.268 &  15.012 &  14.566 &  14.413 &  14.400 \\
        8 & 303110031 & 200.2954 &  5.5717 & 200.2956 &  5.5726 &  14.880 &  15.050 &  14.650 &  15.703 &  14.806 &  14.521 &  14.445 &  14.430 \\
        9 & 303110021 & 200.6808 &  3.6422 & 200.6812 &  3.6434 &  15.300 &  15.350 &  14.800 &  16.131 &  15.034 &  14.691 &  14.548 &  14.519 \\
       10 & 303110026 & 200.7858 &  5.6906 & 200.7862 &  5.6917 &  14.840 &  15.020 &  14.590 &  15.717 &  14.745 &  14.486 &  14.386 &  14.364 \\
       11 & 303110019 & 200.9633 &  2.1536 & 200.9627 &  2.1530 &  14.990 &  14.890 &  14.260 &  15.823 &  14.517 &  14.040 &  13.938 &  13.872 \\
       12 & 303110020 & 201.0854 &  3.4172 & 201.0859 &  3.4179 &  15.210 &  15.180 &  14.680 &  16.037 &  14.906 &  14.531 &  14.426 &  14.441 \\
       13 & 303110015 & 201.5829 &  3.9639 & 201.5831 &  3.9651 &  14.280 &  15.220 &  15.350 &  15.028 & 15.181 &  15.423 &  15.509 &  15.617 \\
       14 & 303110004 & 202.0821 &  3.3917 & 202.0821 &  3.3920 &  14.950 &  14.970 &  14.490 &  15.855 &  14.703 &  14.355 &  14.166 &  14.114 \\
       15 & 164770013 & 219.0267 &  3.6642 & 219.0262 &  3.6641 & $\cdots$ &  14.488 &  14.058 &  15.138 &  14.156 &  13.855 &  13.790 &  13.748 \\
       16 & 164770026 & 219.3088 &  2.7725 & 219.3079 &  2.7723 &  15.062 &  15.064 &  14.714 &  16.060 &  14.993 &  14.747 &  14.699 &  14.660 \\
       17 & 164770031 & 219.7417 &  4.0964 & 219.7411 &  4.0969 &  15.081 &  15.148 &  14.868 &  15.928 &  14.887 &  14.766 &  14.753 &  14.791 \\
       18 & 164770028 & 219.9288 &  2.9500 & 219.9286 &  2.9489 & $\cdots$ &  14.894 &  14.254 &  15.760 &  14.519 &  14.089 &  13.954 &  13.932 \\
       19 & 303170034 & 220.3283 &  4.0239 & 220.3291 &  4.0244 &  14.802 &  14.874 &  14.284 &  15.631 &  14.566 &  14.255 &  14.179 &  14.158 \\
       20 & 164770057 & 220.8487 &  4.8675 & 220.8486 &  4.8677 &  15.232 &  15.096 &  14.966 &  16.155 &  14.894 &  15.037 &  15.172 &  15.251 \\
       21 & 164770054 & 221.0037 &  3.9461 & 221.0035 &  3.9455 & $\cdots$ &  15.146 &  14.426 &  16.237 &  14.706 &  14.152 &  14.000 &  13.902 \\
       22 & 169810101 & 221.1496 & -0.9181 & 221.1499 & -0.9187 &  14.872 &  14.974 &  14.464 &  15.553 &  14.561 &  14.238 &  14.124 &  14.100 \\
       23 & 169810110 & 221.5996 & -0.1158 & 221.5997 & -0.1158 &  15.622 &  {\bf 15.494} &  15.184 &  15.856 & {\bf 14.711}  & 14.726  &  14.770  &  14.834 \\
       24 & 164770069 & 221.6121 &  3.6933 & 221.6126 &  3.6946 & $\cdots$ &  15.506 &  14.866 &  16.372 &  15.128 &  14.681 &  14.601 &  14.546 \\
       25 & 303170014 & 221.7533 &  4.9250 & 221.7531 &  4.9256 &  14.392 &  14.566 &  14.116 &  15.253 &  14.371 &  14.122 &  14.066 &  14.050 \\
       26 & 169810119 & 222.4554 &  2.0433 & 222.4559 &  2.0423 &  15.232 &  15.104 &  14.524 &  15.888 &  14.674 &  14.239 &  14.152 &  14.053 \\
       27 & 303010160 & 222.8217 &  0.7903 & 222.8224 &  0.7917 &  15.082 &  15.114 &  14.774 &  15.809 &  14.860 &  14.772 &  14.769 &  14.801 \\
       28 & 164770095 & 223.0192 &  3.3558 & 223.0181 &  3.3569 &  13.952 &  13.994 &  13.634 &  14.733 &  13.708 &  13.514 &  13.489 &  13.562 \\
       29 & 303250103 & 223.2250 &  3.0119 & 223.2244 &  3.0126 &  14.472 &  14.654 &  14.204 &  15.277 &  14.400 &  14.101 &  13.983 &  13.969 \\
       30 & 303010147 & 223.4338 &  1.6775 & 223.4338 &  1.6780 &  15.033 &  14.972 &  14.372 &  15.811 &  14.639 &  14.199 &  14.092 &  14.002 \\
       31 & 303010122 & 224.1288 &  1.5625 & 224.1288 &  1.5633 & $\cdots$ &  14.482 &  14.362 &  15.393 &  14.210 &  14.277 &  14.452 &  14.522 \\
       32 & 164720014 & 224.1600 &  1.0669 & 224.1595 &  1.0662 & $\cdots$ &  14.894 &  14.364 &  15.593 &  14.477 &  14.148 &  14.032 &  14.013 \\
       33 & 303010119 & 224.1821 &  2.0694 & 224.1822 &  2.0694 &  15.573 &  15.652 &  15.112 &  16.347 &  15.345 &  14.988 &  14.846 &  14.804 \\
       34 & 303250074 & 224.3604 &  2.9856 & 224.3599 &  2.9851 &  14.603 &  14.792 &  14.422 &  15.447 &  14.543 &  14.297 &  14.202 &  14.214 \\
       35 & 303010098 & 224.7829 & -0.2539 & 224.7820 & -0.2537 &  14.923 &  14.952 &  14.522 &  15.591 &  14.631 &  14.361 &  14.338 &  14.371 \\
       36 & 303010093 & 225.2812 &  1.9531 & 225.2810 &  1.9540 &  15.422 &  15.564 &  15.064 &  16.214 &  15.255 &  14.907 &  14.792 &  14.761 \\
       37 & 164720041 & 225.4275 &  1.2161 & 225.4274 &  1.2165 & $\cdots$&  14.172 &  13.802 &  14.928 &  13.828 &  13.637 &  13.597 &  13.646 \\
       38 & 303010084 & 225.5146 &  1.0722 & 225.5137 &  1.0718 &  14.523 &  14.422 &  13.792 &  15.354 &  14.074 &  13.591 &  13.429 &  13.431 \\
       39 & 303010009 & 227.4396 & -0.0919 & 227.4394 & -0.0915 &  14.893 &  14.932 &  14.322 &  15.604 &  14.534 &  14.040 &  13.860 &  13.756 \\
       40 & 228940007 & 353.0421 &  0.0939 & 353.0421 &  0.0937 &  13.840 &  13.699 &  13.534 &  14.669 &  13.511 &  13.584 &  13.688 &  13.902 \\
       41 & 228940006 & 353.0750 & -0.0731 & 353.0752 & -0.0736 &  15.663 &  15.506 &  15.444 &  16.574 &  15.368 &  15.507 &  15.644 &  15.707 \\
       42 & 228940004 & 353.1646 & -0.9369 & 353.1648 & -0.9368 &  14.260 &  14.456 &  14.074 &  15.028 &  14.232 &  13.980 &  13.903 &  13.837 \\
       43 & 228940009 & 353.3700 &  0.9594 & 353.3702 &  0.9591 &  14.407 &  14.538 &  14.044 &  15.276 &  14.270 &  13.920 &  13.730 &  13.703 \\
       44 & 228940005 & 353.4475 & -0.5358 & 353.4482 & -0.5360 &  14.698 &  14.859 &  14.414 &  15.528 &  14.617 &  14.290 &  14.154 &  14.132 \\
       45 & 228940003 & 353.4846 & -1.2039 & 353.4847 & -1.2041 &  14.806 &  14.646 &  14.544 &  15.755 &  14.610 &  14.664 &  14.744 &  14.811 \\
       46 & 228940014 & 353.9450 &  0.8819 & 353.9454 &  0.8817 &  14.645 &  14.526 &  14.434 &  15.621 &  14.362 &  14.484 &  14.578 &  14.661 \\
       47 & 228940017 & 354.0967 &  0.0700 & 354.0973 &  0.0702 &  14.321 &  14.487 &  14.034 &  15.116 &  14.283 &  13.928 &  13.752 &  13.841 \\
       48 & 228940020 & 354.5200 & -0.4206 & 354.5199 & -0.4203 &  15.000 &  15.069 &  14.764 &  15.911 &  14.852 &  14.712 &  14.683 &  14.731 \\
       49 & 228940019 & 354.8287 &  0.0617 & 354.8294 &  0.0616 &  14.014 &  14.271 &  13.856 &  14.794 &  14.044 &  13.769 &  13.616 &  13.675 \\
       50 & 228940030 & 354.9933 & -0.8097 & 354.9939 & -0.8094 &  15.034 &  14.974 &  14.586 &  15.908 &  14.767 &  14.530 &  14.428 &  14.432 \\
       51 & 228940031 & 355.0013 & -0.5106 & 355.0020 & -0.5109 &  14.290 &  14.162 &  14.116 &  15.255 &  14.073 &  14.229 &  14.352 &  14.506 \\
       52 & 228940033 & 355.0225 & -0.3222 & 355.0226 & -0.3219 &  15.183 &  15.032 &  14.926 &  16.069 &  14.890 &  15.046 &  15.161 &  15.198 \\
       53 & 228940034 & 355.4050 &  0.0236 & 355.4052 &  0.0231 &  14.988 &  14.865 &  14.696 &  15.889 &  14.634 &  14.679 &  14.748 &  14.886 \\
       54 & 228940036 & 355.5304 &  0.5481 & 355.5305 &  0.5480 &  15.011 &  14.955 &  14.696 &  15.840 &  14.767 &  14.662 &  14.638 &  14.606 \\
       55 & 228940029 & 355.5604 & -0.8592 & 355.5609 & -0.8591 &  14.453 &  14.533 &  14.274 &  15.407 &  14.348 &  14.227 &  14.275 &  14.311 \\
       56 & 228940028 & 355.8275 & -1.1953 & 355.8280 & -1.1952 &  14.862 &  15.064 &  14.634 &  15.669 &  14.713 &  14.505 &  14.332 &  14.364 \\
       57 & 228940046 & 356.1296 & -0.4364 & 356.1294 & -0.4363 &  15.252 &  15.214 &  14.766 &  16.197 &  14.945 &  14.607 &  14.521 &  14.477 \\
       58 & 228940054 & 357.1179 &  0.4003 & 357.1192 &  0.3994 &  14.361 &  14.486 &  14.116 &  15.246 &  14.182 &  13.980 &  13.981 &  13.906 \\

\enddata
\end{deluxetable}


\begin{thebibliography}{}

\bibitem[Bilir, Karaali \&Tuncel(2005)]{bkt05} Bilir, S., Karaali, S., \& Tuncel, S. 2005, Astronomische Nachrichten, {\bf 326}, 321
\bibitem[Cardelli, Claydon, \& Mathis(1989)]{ccm89} Cardelli, J.A., Clayton, G.C., \& Mathis, J.S. 1989, \apj, {\bf 345}, 245
\bibitem[Fukugita et al.(1996)]{fetal96} Fukugita, M. et al. 1996, \aj, {\bf 111}, 1748
\bibitem[Gunn et al.(1998)]{getal98} Gunn, J. E., et al. 1998, \aj, {\bf 116}, 3040
\bibitem[Hogg et al.(2001)]{hetal01} Hogg, D. W., et al. 2001, \aj, {\bf 122}, 2129
\bibitem[Jester et al.(2005)]{jetal05} Jester, S. et al. 2005, \aj, {\bf 130}, 873
\bibitem[Karaali, Bilir \& Tuncel(2005)]{kbt05} Karaali, S., Bilir, S., \& Tuncel, S. 2005, Proc. Astr. Soc. Australia, {\bf 22}, 24
\bibitem[Mihalas \& Binney(1981)]{mb81} Mihalas, D. \& Binney, J. 1981, {\it Galactic Astronomy, Structure and Kinematics}, Second Edition, published by W.H. Freeman and Company 
\bibitem[Pier et al.(2003)]{petal03} Pier, J. R., et al. 2003, \aj,  {\bf}, 1559
\bibitem[Preston, Schectman \& Beers(1991)]{psb91} Preston, G. W., Shectman, S. A., \& Beers, T. C. 1991, \apjs, {\bf 76}, 1001
\bibitem[Schlegel, Finkbeiner \& Davis(1998)]{fd98} Schlegel, D.J., Finkbeiner, D.P., \& Davis, M. 1998, \apj, {\bf 500}, 525
\bibitem[Smith et al.(2002)]{setal02} Smith, J. A. et al. 2002, \aj, {\bf 123}, 2121
\bibitem[West, Walkowicz \& Hawley(2005)]{wwh05} West, A. A., Walkowicz, L. M., \& Hawley, S. L. 2005, \pasp, {\bf 117}, 706
\bibitem[York et al.(2000)]{yetal00} York, D. G. et al. 2000, \aj, {\bf 120}, 1579
\end{thebibliography}
\end{document}